\documentclass[12pt]{article}
\usepackage{graphics,epsfig}

\begin{document}

\title{Getting bounds on the mixing angles for a non-sequential bottom quark}
\author{R. Martinez, J-Alexis Rodriguez and M. Vargas \\
Dpto. de Fisica, Universidad Nacional de Colombia\\
Bogota, Colombia}
\date{}
\maketitle

\begin{abstract}
We analyze the $Zf\overline{f}$ vertex in the framework of models that add a new bottom
quark in a nonsequential way and we evaluate the tree level contribution to
the LEP/SLC observables $\Gamma _{Z}$, $R_{l}$ and $R_{b}$. We obtain bounds
for the mixing angles from the experimentally allowed contour regions of the
parameters $\Lambda _{L,R}$ introduced here. In order to get a more
restrictive region, we consider the experimental results for $B\rightarrow
\nu \overline{\nu} X$ as well.
\end{abstract}


\section{Introduction}

The comparison of theoretical predictions with experimental data has
confirmed the validity of the Standard Model (SM) in an impressive way. The
quantum effects of the SM have been established at the $1\sigma $ level, and
the direct and indirect determinations of the top quark mass are compatible
with each other. In spite of this success, the conceptual situation with the
SM is not completely satisfactory for a number of deficiencies. Some of them
are the large number of free parameters and the hierarchical fermion masses.

The SM contains three generations of quarks in irreducible representations
of the gauge symmetry group $SU(3)_{C}\times SU(2)_{L}\times U(1)_{Y}$. The
possibility of extending them has been studied in different frameworks \cite
{Barger}-\cite{Silva} which are based either on a fourth generation
sequential family, or on non-sequential fermions, regularly called exotic
representations because they are different from those of the SM.

These unusual representations emerge in other theories, like the $E_{6}$
model where a singlet bottom type quark appears in the fundamental
representation \cite{Nir}; also, top-like singlets have been suggested in
supersymmetric gauge theories\cite{Branco}. The principal feature of a model
which extends the quark sector with an exotic fermion is that there are new
quark mixing phases in addition to the single phase of the SM. Therefore, in
this kind of models $Z$ boson mediated FCNC's arise at tree level. This fact
can affect the mixing mechanism in the neutral $B$-system \cite{Barger}-\cite
{Gronau}.

The possibility of indirect consequences of singlet quark mixing for FCNC
and CP violation has been used to get bounds on the flavor changing
couplings. Heavy meson decays like $B^{0}$ and $D^{0}\;\rightarrow 
\mu^{+}\mu ^{-}$ \cite{Barger}, \cite{Dib}, rare decays $b\rightarrow s\gamma $ 
\cite{Barger}, \cite{Nir}, \cite{Dib}, measurements like $K_{L}\rightarrow
\mu^{+}\mu ^{-}$, $B\rightarrow \mu^{+}\mu ^{-} X$, $B\rightarrow \nu \overline{\nu} X$, 
$K$ meson physics \cite{Barger},\cite{Nir}, \cite{Nardi}, or even $Z\rightarrow 
l\overline{l}$, $l\rightarrow ll\overline{l}$ \cite{Cotti} have been considered 
for this purpose.

In the last years, the LEP and SLC colliders have brought to completion a
remarkable experimental program by collecting an enormous amount of
electroweak precision data on the $Z$ resonance. This activity, together
with the theoretical efforts to provide accurate SM predictions have formed
the apparatus of electroweak precision tests \cite{Altarelli}. We are
interested in using the electroweak precision test quantities in order to
get bounds on the mixing angles for additional fermions in exotic
representations. Specifically, we want to consider models that include a new
quark with charge $-1/3$ which is mixed with the SM bottom quark. This kind
of new physics was taken into account by Bamert, et. al. \cite{Bamert}
during the discrepancy between experiment and SM theory in the $R_{b}$
ratio. They analyzed a broad class of models in order to explain the
discrepancy, and they considered those models in which new $Zb\overline{b}$ couplings
arise at tree level through $Z$ or $b$ quark mixing with new particles.

Our presentation is based on the parametrization of the $Zf\overline{f}$ vertex in an
independent model formulation. Therefore these results can be used for
different quark representations like singlet down quark, vector doublets
model, mirror fermions and self conjugated triplets, etc. The
parametrization of the vertex in a general way has been reviewed by Barger
et. al. \cite{Barger}, \cite{Nardi} as well as Cotti and Zepeda \cite{Cotti}. 
The LEP precision test parameters that we use are the total $Z$ width $%
\Gamma _{Z}$, $R_{l}$ and $R_{b}$.

The procedure to get bounds on the mixing angles is the following. First, we
analyze the $Zf\overline{f}$ vertex as obtained after a rotation of a general quark
multiplet (common charge) into mass eigenstates. In particular, we write
down the neutral current terms for the bottom quarks, which are assumed to
be mixed. With these expressions we can evaluate the tree level contribution
to the process $Z\rightarrow b\overline{b}$; we enclose this new contribution within
the coupling constants $g_{V}$ (vectorial) and $g_{A}$ (axial). We then
write down $\Gamma _{Z}$, $R_{l}$ and $R_{b}$ including the new
contributions, and we obtain bounds on the new parameters by using the
experimental values from LEP and SLC \cite{EuropPhysJour}. Finally, we do a $%
\chi ^{2}$ analysis and find the allowed region in the plane of the new
parameters $\Lambda _{L}$ and $\Lambda _{R}$ introduced. We also use the
result obtained by Grossman et. al., involving $B\rightarrow \nu \overline{\nu} X_{s}$%
\cite{Nardi}, in order to narrow down the bounds in the contour plots.

\section{Precision test parameters}

To restrict new physics, we will use parameters measured at the $Z$ pole.
These parameters are the total decay width of the $Z$ boson $\Gamma _{Z}$,
the fractions $R_{b}=\Gamma \left( Z\rightarrow b\overline{b}\right) /\Gamma
\left( Z\rightarrow hadrons\right) $ and $R_{l}=\Gamma \left( Z\rightarrow
hadrons\right) $ $/$ $\Gamma \left( Z\rightarrow l\overline{l}\right) $ 
\cite{Altarelli}, \cite{EuropPhysJour}. Considering the new physics (NP) and the
SM couplings, we can write 
\begin{eqnarray}
\Gamma \left( Z\rightarrow b\overline{b}\right)  &=&\frac{G_{F}m_{Z}^{3}}{6%
\sqrt{2}\pi }\left[ \frac{3\beta -\beta ^{3}}{2}\left(
g_{V}^{SM}+g_{V}^{NP}\right) ^{2}+\beta ^{3}\left(
g_{A}^{SM}+g_{A}^{NP}\right) ^{2}\right]   \nonumber \\
&\times &N_{C}R_{QCD+QED}
\end{eqnarray}
where $N_{C}$ is the number of colors, $R_{QCD+QED}$ are the QCD and QED
corrections, and $\beta =\sqrt{1-4\frac{m_{b}^{2}}{m_{Z}^{2}}}$ is the
kinematic factor \cite{Altarelli} with $m_{b}=4.7$ GeV. We also are taking
into account the oblique and vertex contributions to $g_{V,A}^{SM}$ giving
by the top quark and Higgs boson. For our purpose, It is convenient to
separate the SM and NP contributions as follows: 
\begin{equation}
\Gamma \left( Z\rightarrow b\overline{b}\right) =\Gamma _{b}^{SM}\left(
1+\delta _{b}^{NP}\right) .
\end{equation}
The symbol $\delta _{b}^{NP}$ is given by: 
\begin{equation}
\delta _{b}^{NP}=\frac{\left( 3-\beta ^{2}\right) \left[ \left(
g_{V}^{NP}\right) ^{2}+2g_{V}^{NP}g_{V}^{SM}\right] +2\beta ^{2}\left[
\left( g_{A}^{NP}\right) ^{2}+2g_{A}^{NP}g_{A}^{SM}\right] }{\left( 3-\beta
^{2}\right) \left( g_{V}^{SM}\right) ^{2}+2\beta ^{2}\left(
g_{A}^{SM}\right) ^{2}}.
\end{equation}
This equation could be written using the new physics parameters $\Lambda
_{L\left( R\right) }$ that were introduced in the eq. (\ref{NPparam}), through
the relationships $g_{A}^{NP}=\Lambda _{L}-\Lambda _{R}$ and $%
g_{V}^{NP}=\Lambda _{R}+\Lambda _{L}$

Similarly, the $Z$ decay into hadrons after considering the NP, can be
written as: 
\begin{eqnarray}
\Gamma \left( Z\rightarrow hadrons\right) &=&2\Gamma _{u}^{SM}+2\Gamma
_{d}^{SM}+\Gamma _{b}  \nonumber \\
&=&\Gamma _{had}^{SM}\left( 1+\frac{\Gamma _{b}^{SM}}{\Gamma _{had}^{SM}}
\delta _{b}^{NP}\right) .  \label{hadron}
\end{eqnarray}
Here, only $\Gamma_b$ gets NP corrections because only the SM bottom mixes
with the exotic quark. Therefore, the $Z$ partial decay into $d$ and $s$
quarks remains unchanged.

On the other hand, $\Gamma _{Z}$ is equal to 
\begin{equation}
\Gamma _{Z}=3\Gamma \left( Z\rightarrow \nu \overline{\nu }\right) +3\Gamma
\left( Z\rightarrow l\overline{l}\right) +\Gamma \left( Z\rightarrow
hadrons\right)
\end{equation}
which again is re-written, with the eq.(\ref{hadron}), as follows: 
\begin{equation}
\Gamma _{Z}=\Gamma _{Z}^{SM}\left( 1+\frac{\Gamma _{b}^{SM}}{\Gamma _{Z}^{SM}%
}\delta _{b}^{NP}\right) .
\end{equation}

Using the above equations, for $R_l$ and $R_b$ we obtain the following
expressions: 
\begin{eqnarray}
R_{l} &=&R_{l}^{SM}\left( 1+R_{b}^{SM}\delta _{b}^{NP}\right) ,  \nonumber \\
R_{b} &=&R_{b}^{SM}\left[ 1+\delta _{b}^{NP}\left( 1-R_{b}^{SM}\right) %
\right] .
\end{eqnarray}
In a general way, $R_{b}$ is mainly a measure of $\left| g_{L}^{b}\right|
^{2}+\left| g_{R}^{b}\right| ^{2}$; therefore, the fraction $R_{b}$ is very
sensitive to anomalous couplings of the $b$ quark.

\section{The model}

Following closely the notation of ref. \cite{Cotti}, if we have a multiplet $%
\Psi _{a=L,R}^{O}$ with $n_{a}$ ordinary fermions and $m_{a}$ exotic
fermions with the same electric charge $q$: 
\begin{equation}
\Psi _{a}^{O}=U_{a}\Psi _{a},\;\;\;\;\Psi _{a}=\left( 
\begin{array}{c}
\Psi _{l} \\ 
\Psi _{h}
\end{array}
\right) _{a}
\end{equation}
where $U_{a}$ is the unitary matrix that rotates the mass eigenstate $\Psi
_{a}$ into the interaction eigenstate $\Psi _{a}^{O}$. $\Psi_{l(h)}$ means
ordinary or light (exotic or heavy) fermions. $U_{a}$ can be further the
composed as follows\cite{Cotti}: 
\begin{equation}
U_{a}=\left( 
\begin{array}{cc}
A & E \\ 
F & G
\end{array}
\right) _{a}
\end{equation}
where 
\begin{equation}
\left( U^{+}U\right) _{a}=\left( 
\begin{array}{cc}
A^{+}A+F^{+}F & A^{+}E+F^{+}G \\ 
E^{+}A+G^{+}F & E^{+}E+G^{+}G
\end{array}
\right) _{a}=\left( 
\begin{array}{cc}
1 & 0 \\ 
0 & 1
\end{array}
\right) \;\;\; .  \label{equation3}
\end{equation}

If we suppose that the up quark sector of the SM is diagonal and that there
are no exotic quarks, then $A_{L}$ corresponds to the classical
Kobayashi-Maskawa matrix. In the SM this matrix is unitarity, whereas in our
model it is not: 
\begin{equation}
\left( A^{+}A\right) _{L}=I-\left( F^{+}F\right) _{L} .
\end{equation}
$F_{L}$ corresponds to the mixing of the ordinary-exotic quarks. As
mentioned, $A_L$ is not quite unitary and the factor $\left( F^{+}F\right)
_{L}$ indicates Flavor Changing transitions in the light-light sector.

The neutral current Lagrangian for the multiplet $\Psi $ is given by 
\begin{eqnarray}
-{\cal L}^{NC} &=&\frac{e}{c_{w}s_{w}}\sum_{a=L,R}\overline{\Psi ^{O}}%
_{a}\gamma ^{\mu }D_{a}\Psi _{a}^{O}Z_{\mu }^{0},  \nonumber \\
&=&\frac{e}{c_{w}s_{w}}\sum_{a=L,R}\overline{\Psi }_{a}\gamma ^{\mu
}U_{a}^{+}D_{a}U_{a}\Psi _{a}Z_{\mu }^{0}  \label{lagrangiano}
\end{eqnarray}
where $s_{w}=\sin \theta _{w}$ and $D_{a}$ are diagonal matrices which
contain the couplings of the neutral $Z^0$ gauge boson to the matter fields;
they have the form: 
\begin{eqnarray}
D_{a} &=&\left( T_{3}- Q s_{w}^{2}\right) _{a},  \nonumber \\
&=&\left( 
\begin{array}{cc}
t_{30}-qs_{w}^{2} & 0 \\ 
0 & t_{3E}-qs_{w}^{2}
\end{array}
\right) _{a}
\end{eqnarray}
where $T_{3a}$ and $Q$ are the matrices of the isospin charges and the electric charge, 
respectively. $t_{30}$ and $t_{3E}$ are the standard and exotic weak isospin 3rd
componend of the multiplets. Using the unitarity relations of the $U_a$
matrix from the eq. (\ref{equation3}), the product $\left( U^{+}DU\right)
_{a=L,R}$ in eq. (\ref{lagrangiano}) can be written as: 
\begin{eqnarray}
\left( U^{+}DU\right) _{L} &=&\left( 
\begin{array}{cc}
F^{+}F & -A^{+}E \\ 
-E^{+}A & -E^{+}E
\end{array}
\right) _{L}\left( t_{3E}-t_{30}\right) _{L}+T_{3L} - Q s_{w}^{2}, 
\nonumber \\
\left( U^{+}DU\right) _{R} &=&\left( 
\begin{array}{cc}
F^{+}F & F^{+}G \\ 
G^{+}F & G^{+}G
\end{array}
\right) _{R}t_{3ER}- Q s_{w}^{2}
\end{eqnarray}

The neutral current Lagrangian in the light-light sector can be written as 
\cite{Cotti}: 
\begin{equation}
{\cal L}^{NC}=-\frac{e}{c_{w}s_{w}}\sum_{a=L,R}\overline{\Psi }_{l,a}\gamma
^{\mu }K_{a}\Psi _{l,a}Z_{\mu }^{0}  \label{NClagrang}
\end{equation}
where 
\begin{eqnarray}
K_{L} &=&\left( F^{+}F\right) _{L}\left( t_{3EL}-t_{30L}\right) +I_{3\times
3}\left( t_{30L}-qs_{w}^{2}\right) ,  \nonumber \\
K_{R} &=&\left( F^{+}F\right) _{R}t_{3ER}-I_{3\times 3}\;qs_{w}^{2}.
\end{eqnarray}
For the SM with three generations, $K_{L,R}$ are $3\times 3$ matrices. They
can be produced FC transitions at the tree level depending of $F_{L,R}$
entries which are the mixing angles of the ordinary and exotic fermions.

In this work, we only consider one $bottom$ exotic quark (i.e. not mixing
with $d$ and $s$). Then, $U_{a}$ and the $F^{\dagger}F$ product become: 
\begin{eqnarray}
U_{a}=\left( 
\begin{array}{cccc}
&  &  & 0 \\ 
& {\huge A} &  & 0 \\ 
&  &  & -s_a \\ 
0 & 0 & s_a & c_a
\end{array}
\right) \;\; , \;\; (F^{\dagger}F)_a=\left( 
\begin{array}{ccc}
0 & 0 & 0 \\ 
0 & 0 & 0 \\ 
0 & 0 & \sin^2\theta_a
\end{array}
\right) .\;\;
\end{eqnarray}
where $\sin\theta_{L,R}$ represent the mixing between bottom quark with the
exotic ones. Therefore, the coupling $bbZ$ gets modified by the $%
\Lambda_{L,R}$ factors: 
\begin{eqnarray}
K_{L}^b &=&\Lambda _{L}+t_{30L}-qs_{w}^{2},  \nonumber \\
&=&\sin^2\theta_L \left(t_{3EL}+\frac{1}{2}\right)+\left(-\frac{1}{2}+ \frac{%
1}{3}s_{w}^{2}\right),  \nonumber \\
K_{R}^b &=&\Lambda_{R}-qs_{w}^{2},  \nonumber \\
&=&\sin^2\theta_R \; t_{3ER}+ \left(\frac{1}{3} s_{w}^{2}\right).
\label{NPparam}
\end{eqnarray}

\section{Results}

With the expressions for $\Gamma _{Z}$, $R_{l}$ and $R_{b}$ in terms of the
new physics contribution in section 3, and with the experimental data from
LEP we get bounds on the parameters $\Lambda _{L,R}$ introduced in eq. (\ref
{NPparam}). The experimental data that we used for the LEP parameters, as
well as their SM values are in Table 1 \cite{Altarelli}, \cite{EuropPhysJour}.

We do a $\chi ^{2}$ fit of the observables $\Gamma _{Z}$, $%
R_{l}$ and $R_{b}$, and then we proceed to obtain bounds on the parameters $%
\Lambda _{L,R}$ by taking on values in the best region allowed for them at $95\%$ C.L.
This region is displayed in the figure. In order to get a more restrictive
region we use the bound $\left| \Lambda _{L,R}\right| <0.0018$ obtained by
Grossman et.al. \cite{Nardi}, which is represented by straight lines in the
figure. The intersection between the two regions is given by: 
\begin{eqnarray}
-1.094\times 10^{-4} &\leq &\Lambda _{L}\leq 1.086\times 10^{-4},  \nonumber
\\
-1.8\times 10^{-3} &\leq &\Lambda _{R}\leq 1.8\times 10^{-3}.
\end{eqnarray}
We note that for $\Lambda _{L}$ the region is more restrictive than the one
obtained by Grossman et. al. \cite{Nardi}, while the $\Lambda _{R}$
parameter is not modified.

If we consider only mixing between an exotic bottom quark with the third SM
family, independent of any group representations, it is given by a $2\times
2 $ unitary matrix for left- and right-handed fermions. The couplings for
several $SU(2)_{L}$ representations are given in table 2. We can use these
bounds in order to get constraints for the left and right mixing angles of
each model. They are shown in table 3.

Summarizing, we have used the fractions $\Gamma _{Z}$, $R_{l}$ and $R_{b}$
to obtain bounds on the mixing angles of new quark bottom-type
representations with the SM bottom quark. Taking into account the results of
Grossman et. al.\cite{Nardi}, we have gotten the allowed intervals $%
-1.094\times 10^{-4}\leq \Lambda _{L}\leq 1.086\times 10^{-4}$ and $-1.8\times
10^{-3}\leq \Lambda _{R}\leq 1.8\times 10^{-3}$. Our results reduce the
allowed region for the parameter $\Lambda _{L}$ while the parameter $\Lambda
_{R}$ is not modified with respect to the results obtained by Grossman et.
al.\cite{Nardi}. We may note that the results have been obtained from the
tree level contributions, and we can get bounds on the mass of the new quark
using oblique corrections \cite{dario}.

This work was partially supported by COLCIENCIAS and CONACyT (Mexico). One
of us (M. V.) acknowledges the scholarship by Fundaci\'{o}n MAZDA para el
Arte y la Ciencia..

\newpage

\begin{center}
List of tables
\end{center}

\begin{table}[tbph]
\begin{center}
\begin{tabular}{||c||c|c||}
\hline\hline
& {\bf Experimentals} & {\bf Standard Model} \\ \hline\hline
$\Gamma _{Z}$ & $2.4939\pm 0.0024$ & $2.49582$ \\ \hline
$R_{l}$ & $20.765\pm 0.026$ & $20.7468$ \\ \hline
$R_{b}$ & $0.21656\pm 0.00074$ & $0.215894$ \\ \hline\hline
\end{tabular}
\end{center}
\caption{SM predictions and experimental values measured at LEP for the $%
\Gamma _{Z}$, $R_{l}$ and $R_{b}$}
\end{table}

\begin{table}[tbph]
\begin{center}
\begin{tabular}{||l||l|c|r||}
\hline\hline
$\left( t_{EL}^{3},t_{ER}^{3}\right) $ & ${\bf \Lambda }_{L}$ & ${\bf \
\Lambda }_{R}$ & {\bf Model} \\ \hline\hline
$\left( 0,0\right) $ & $\frac{1}{2}$sin$^{2}\theta _{L}$ & 0 & Vector
singlets \\ \hline
$\left( -\frac{1}{2},-\frac{1}{2}\right) $ & $0$ & $-\frac{1}{2}\sin
^{2}\theta _{R}$ & Vector Doublets \\ \hline
$\left( 0,-\frac{1}{2}\right) $ & $\frac{1}{2}\sin ^{2}\theta _{L}$ & $-%
\frac{1}{2}\sin ^{2}\theta _{R}$ & Mirror fermions \\ \hline
$\left( -1,-1\right) $ & $-\frac{1}{2}\sin ^{2}\theta _{L}$ & $-\sin
^{2}\theta _{R}$ & Self-conjugated triplets \\ \hline
\end{tabular}
\end{center}
\caption{The parameters $\Lambda _{L,R}$ for different representations
according with the quantum numbers in eq.(\ref{NClagrang})}
\end{table}

\begin{table}[tbph]
\begin{center}
\begin{tabular}{||l||l|l||}
\hline\hline
{\bf Model} & $\ \left| \sin \theta _{L}\right| \leq $ & $\ \ \left| \sin
\theta _{R}\right| \leq $ \\ \hline
Vector Singlets & $4.661\times 10^{-2}$ & $\ \ \ 0$ \\ \hline
Vector doublets & $\ \ \ \ \ \ 0$ & $6\times 10^{-2}$ \\ \hline
Mirror fermion & $4.661\times 10^{-2}$ & $6\times 10^{-2}$ \\ \hline
Self-conjugated triplets & $4.679\times 10^{-2}$ & $4.24\times 10^{-2}$ \\ 
\hline\hline
\end{tabular}
\end{center}
\caption{Bounds on the mixing angles for different representations of the
exotic quarks}
\end{table}

\newpage

\begin{figure}[h]
\begin{center}
\includegraphics[angle=-90, width=9cm]{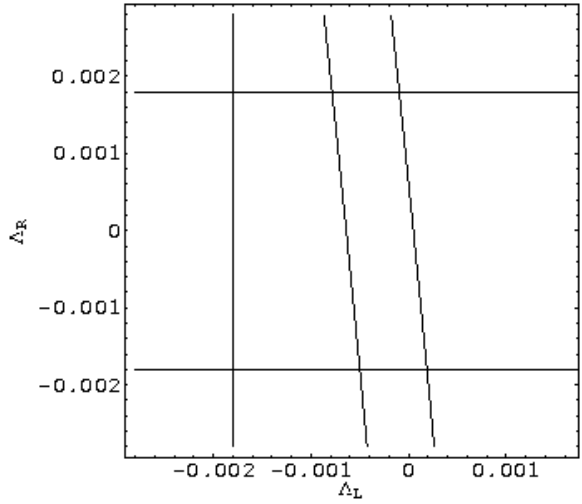}
\end{center}
\caption{Contour plot represents the allowed region for $%
\Lambda_L-\Lambda_R$. The straigth lines are the bounds from $B \to \nu\overline{\nu} X 
$ reported in ref.[4].}
\label{Fig. 1}
\end{figure}

\end{document}